\documentclass[aps,prl,reprint,superscriptaddress]{revtex4-1}

\usepackage{graphicx}% Include figure files
\usepackage{dcolumn}% Align table columns on decimal point
\usepackage{bm}% bold math
\usepackage{xcolor}
\usepackage{ulem}

\newcommand{\beq}{\begin{equation}}
\newcommand{\eeq}{\end{equation}}

\newcommand{\apar}{ A_{\parallel}}

\newcommand{\lapp}{\nabla_{\perp}^2}

\newcommand{\ben}{\begin{eqnarray}}
\newcommand{\een}{\end{eqnarray}}

\newcommand{\rs}{\rho_s}

\begin{document}

\preprint{APS/123-QED}

%\title{Collisionless plasmoid instability (???)}% Force line breaks with \\
%Suggestion:
\title{Marginally Stable Current Sheets in Collisionless Magnetic Reconnection}

\author{C. Granier}\email{camille.granier@oca.eu}
\affiliation{Universit\'e C\^ote d'Azur, CNRS, Observatoire de la C\^ote d'Azur, Laboratoire J. L. Lagrange, Boulevard de l'Observatoire, CS 34229, 06304 Nice Cedex 4, France}
\affiliation{Istituto dei Sistemi Complessi - CNR and Dipartimento di Energia, Politecnico di Torino, Torino 10129, Italy}

\author{D. Borgogno}
\affiliation{Istituto dei Sistemi Complessi - CNR and Dipartimento di Energia, Politecnico di Torino, Torino 10129, Italy}

\author{L. Comisso}
\affiliation{Department of Astronomy and Columbia Astrophysics Laboratory, Columbia University, New York, NY 10027, USA}

\author{D. Grasso}
\affiliation{Istituto dei Sistemi Complessi - CNR and Dipartimento di Energia, Politecnico di Torino, Torino 10129, Italy}

\author{E. Tassi}
\affiliation{Universit\'e C\^ote d'Azur, CNRS, Observatoire de la C\^ote d'Azur, Laboratoire J. L. Lagrange, Boulevard de l'Observatoire, CS 34229, 06304 Nice Cedex 4, France}

\author{R. Numata}%
\affiliation{Graduate School of Information Science, University of Hyogo, Kobe 650-0047, Japan}

%\date{\today}

\begin{abstract}
Non-collisional current sheets that form during the nonlinear development of magnetic reconnection are characterized by a small thickness, of the order of the electron skin depth. They can become unstable to the formation of plasmoids, which allows the magnetic reconnection process to reach high reconnection rates. In this work, we investigate the marginal stability conditions for the development of plasmoids when the forming current sheet is purely collisionless and in the presence of a strong guide field. We analyze the geometry that characterizes the reconnecting current sheet, and what promotes its elongation. Once the reconnecting current sheet is formed, we identify the regimes for which it is plasmoid unstable. Our study shows that plasmoids can be obtained, in this context, from current sheets with an aspect ratio much smaller than in the collisional regime, and that the plasma flow channel of the marginally stable current layers maintains an inverse aspect ratio of $0.1$.
\end{abstract}

\maketitle

Magnetic reconnection is a fundamental plasma process that involves a rapid topological change of the magnetic field leading to an efficient magnetic energy conversion. Magnetic reconnection typically occurs via current sheets (CS), where non-ideal plasma effects become important allowing the change of magnetic field line connectivity \cite{Zwe09,Yam10,Ji22}.

It is well established that the instabilities of thin CS, that lead to the formation of plasmoids, have a fundamental impact on the reconnection rate \cite{Dau09,Bhatta09}. Indeed, even in the resistive magnetohydrodynamics (MHD) framework, the development of plasmoids in the reconnection layer induces a fast magnetic reconnection regime characterized by a reconnection rate that can exceed the estimates based on the Sweet-Parker (SP) theory \cite{Sweet1958,Parker1957} by several orders of magnitude. In collisional CS, it has been shown that plasmoids develop when the Lundquist number $S = L_{\rm{cs}} v_A /(\eta c^2/4\pi)$ exceeds the threshold value $S_\star \sim 10^4$ \cite{Bis96}. Here, the Lundquist number is defined with the length of the CS, $L_{\rm{cs}}$. 
The other quantities are the plasma resistivity $\eta$, the speed of light $c$, and  the Alfv\'en speed $v_A$.  The threshold value on the Lundquist number, $S_\star$, separates the Sweet-Parker regime from the plasmoid-mediated regime of collisional reconnection. In addition, it controls the reconnection rate $R_{\rm{rec}}$ in the plasmoid-mediated regime, $R_{\rm{rec}} \sim S_\star^{-1/2} v_A B_{\rm{up}}$ \cite{Huang10,Uzd10,ComPoP15,CG16}, where $B_{\rm{up}}$ is the reconnecting magnetic field. The extension of the resistive reconnection regime with the inclusion of the ion dynamics associated with the ion sound Larmor radius, $\rs$, or the ion inertial length, $d_i$, complicates the picture. Indeed, when the thickness of the reconnecting CS shrinks below $\rs$ (for reconnection with a guide field) or $d_i$ (for reconnection without guide field), the process becomes even faster and approaches $R_{\rm{rec}} \sim 0.1 v_A B_{\rm{up}}$ \cite{ComBha16}. Given its importance, the transition between the different regimes of reconnection has been thoroughly investigated, and the current understanding of reconnection driven by plasma resistivity has been summarized in the form of parameter space diagrams \cite{Ji11,Dau12,Huang13,Kar13,ComJPP15,Le15,Lou15,Bha18}.

In contrast, the marginal stability of reconnecting CS in the collisionless regime has seen relatively little investigation. This subject was approached in \cite{Ji11}, in which it is argued that, below the scales $d_i$ or $\rho_s$, no plasmoids were formed. Yet, it is acknowledged that reconnection in nature is often driven by collisionless effects beyond the resistive MHD description.

In this Letter, we investigate a phase space described by the two kinetic scales $d_e$  (electron inertial length) and $\rho_s$, compared to the current length $L_{\rm{cs}}$. We show how the aspect ratio of the marginally stable reconnection layer depends on these relevant kinetic scales.  We believe this study might also be useful to support observational and experimental results. In particular, recent observations revealed many reconnection onsets driven by electrons, in the presence of a strong guide field, close to the dayside magnetopause and magnetosheath \cite{Bur16,Pha18}. Moreover, in Ref. \cite{Pha18}, current sheets having a thickness of the order of the electron inertial length were identified.  A study also gave direct experimental proof of plasmoid formation at the X point and at the electron scale in a regime where no plasmoids were predicted by the theory \cite{Ols16}.   \\

We assume a plasma immersed in a strong (guide) magnetic field of amplitude $B_{0}$, resulting in low plasma $\beta$ (the ratio of plasma pressure to magnetic pressure). 
In order to reduce the problem to a few essential ingredients, in our analysis we adopt a simple two-fluid model that retains electron inertia effects, as well as ion sound Larmor radius effects. Specifically, the equations governing the plasma dynamics are \citep[e.g.][]{Caf98}  
\begin{equation} \label{fluid1}
  \frac{\partial n_e}{\partial t} + [\phi, n_e]  = [ \apar, u_e] \, ,
\end{equation} 
\begin{equation} \label{fluid2}
\frac{\partial}{\partial t} \left( \apar - d_e^2 u_e\right) + \left[\phi , \apar - d_e^2  u_e\right]  = \rho_s^2[n_e, \apar] \, ,
\end{equation} 
where $\apar$ and $\phi$ are the magnetic and electrostatic potential, $n_e=\lapp \phi$ is the electron density perturbation, and $u_e=\lapp \apar$ is the parallel electron velocity, also proportional to the current density. The variables are normalized as $\{t,x,\apar, \phi\} = \{  {v_A}\hat{t}/L, {\hat{x}}/{L}, \hat{A}_\parallel/(L B_0), c\hat{\phi}/(v_A L B_0)\} $, where the caret ($\, \hat{} \,$) indicates dimensional quantities and  with $L$ the characteristic equilibrium scale length set by the equilibrium magnetic field.
The normalized  magnetic field and the perpendicular $\bm{E}\times\bm{B}$ velocity are related to $\apar$ and $\phi$ as ${\bm{B}} =  {\bm{\hat z}} + \nabla \apar \times  {\bm{\hat z}}$ and ${\bm{u}}_{\perp}=  {\bm{\hat z}} \times \nabla \phi$, respectively. 
The parameters are $d_e=\sqrt{m_e c^2/{4 \pi e^2 n_0}}/L$ and $\rs=\sqrt{{T_{e} m_i c^2}/{e^2 B_{0}^2}}/L$, corresponding to the normalized electron skin depth and ion sound Larmor radius, respectively. Ions are assumed to be cold. In Eqs. (\ref{fluid1})-(\ref{fluid2}), $[f,g]=\partial_x f \partial_y g - \partial_y f \partial_x g$.

In order to analyze the marginal stability conditions of the plasmoid instability in the collisionless regime, we conducted a large number ($\sim 30$) of numerical simulations of the system of Eqs. (\ref{fluid1})-(\ref{fluid2}). The numerical solver is pseudo-spectral and the advancement in time is done through a third order Adams–Bashforth scheme. We considered a periodic 2D domain $2 L_x \times 2 L_y$, resolved with a number of grid points up to $2000 \times 2400$. On the other hand, in the figures presented in the paper, only a part of the computational domain is shown.
We set up an initial equilibrium given by $\phi^{(0)}(x) = 0$, $\apar^{(0)} (x)=  1/\cosh^2 \left( x  \right)$. The tearing stability parameter for this equilibrium is $ \Delta'_{\rm{box,m}}=  2 [ (5- k_y^2) ( k_y^2+3)]/[ k_y^2 ( k_y^2+4)^{1/2}]$. This equilibrium is tearing unstable if $\Delta_{\rm{box,m}}'>0$, thus for a wavenumber $k_y = \pi m/  L_y < \sqrt{5}$. We will always refer to $\Delta'_{\rm{box}}$ as being associated to the mode $m=1$, and we change its value by taking different box lengths along the $y$ direction. With this set up, one or several tearing modes are initially unstable. The dominant mode generates two magnetic islands separated by a reconnection point (X-point). During the nonlinear phase, a slowly thinning CS forms self-consistently at the X-point location. 
The evolution of this current sheet will be decisive for the formation of plasmoids. 

% HERE STARTS THE DESCRIPTION OF THE RESULTS 

In the following, we characterize this reconnecting CS according to the parameters $d_e$, $\rho_s$, and $\Delta'_{\rm{box}}$. \\ 
Specifically, we measured the length and the width of the CS at a time $t$ just before the plasmoid onset. We define the measure $L_{\rm{cs}}$, such that, taking the variation from the highest current position $u_e|_X$ ($u_e$ evaluated at the $X$-point), the standard deviation of the current distribution from $y=0$ to $y=L_{\rm{cs}}/2$ equals unity, I.e. $\sqrt{{\sum^N_{i=1} \left[u_e|_X - u_e(0, i \Delta y ,t)\right]^2}/{N}}=1$,  where $\Delta y$ is the distance between two grid-points along $y$ and $N$ indicates the number of points from $y=0$ to $y=L_{\rm{cs}}/2$. This method makes it possible to account for the decrease of the current intensity along the layer. Once $L_{\rm{cs}}$ is identified, the half width of the CS, which we denote by $\delta_{\rm{cs}}/2$, corresponds to the distance, along $x$, between $u_e|_X$ and the position where the current reaches the value $u_e(\delta_{\rm{cs}}/2,0)= u_e(0, L_{\rm{cs}}/2)$. We also measure the width and length of the outflow velocity channel coming out from the end of the CS. The length $L_{\rm{outf}}$ corresponds to the distance between the upward and downward peaks in the distribution of $u_y=\partial_x \phi$. while the width  $\delta_{\rm{outf}}$ is also measured with the standard deviation method. The aspect ratios $A_{\rm{cs}}=L_{\rm{cs}}/\delta_{\rm{cs}}$ and $A_{\rm{outf}}=L_{\rm{outf}}/\delta_{\rm{outf}}$ are also reported.

\begin{figure}
    \centering
    % to crop the figures: trim=left bottom right top
\includegraphics[trim=0 0 0 0, width=8.7cm]{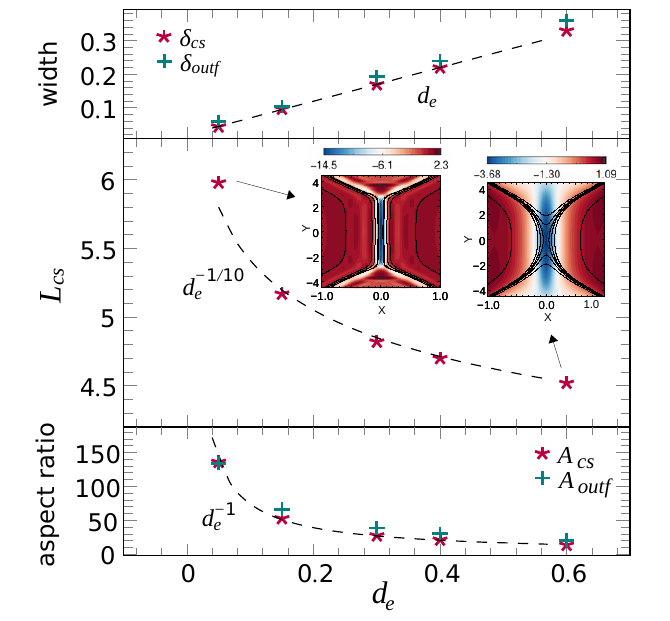}
\vspace{-0.55cm}
    \caption{Characteristics of the reconnecting CS as a function of $d_e$ for fixed $\rho_s=0$ and $\Delta'_{\rm{box}} = 60$. The CS is unstable to the formation of plasmoids in all five cases. For $d_e=0.05$ and $d_e=0.6$, we show the color maps of $u_e$ with isolines of $\apar$ in black.}
    \label{fig:1}
\end{figure}

\begin{figure} 
    \centering
    % to crop the figures: trim=left bottom right top
\includegraphics[trim=0 0 0 0, width=9.cm]{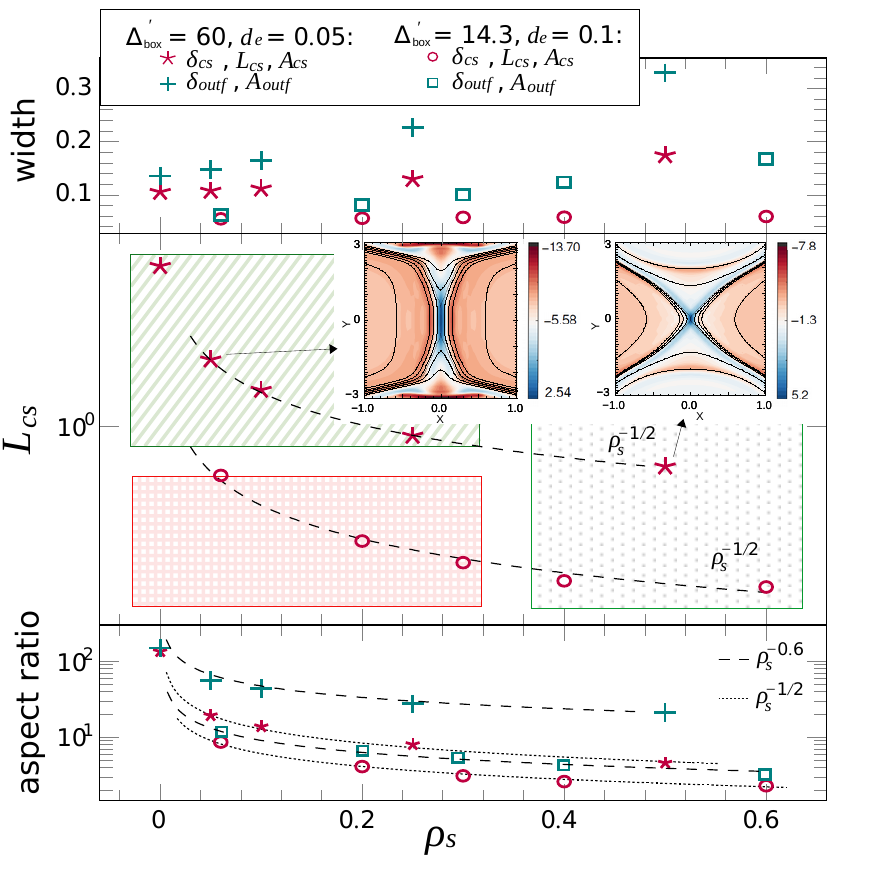}
\vspace{-0.55cm}
    \caption{ Same measures as Fig. \ref{fig:1} but for fixed $d_e$ and varying $\rho_s$. For $\Delta'_{\rm{box}}=60$ and $d_e=0.05$, all five cases are plasmoid unstable. For $\Delta'_{\rm{box}}=14.3$ and $d_e=0.1$,  plasmoids grow only when $\rho_s \geq 0.4$. For $\rho_s=0.05$ and $\rho_s=0.5$, we show the color maps of $u_e$ with isolines of $\apar$ in black.}
    \label{fig:2}
\end{figure}

We first focus on the limit $\rho_s=0$ shown in Fig. \ref{fig:1}. As discussed in Ref. \cite{OP93}, in this limit $\delta_{\rm{cs}}\propto d_e$. To better show the geometry, the colored contour maps of $u_e$ with superimposed contour lines of $\apar$ in black, are shown for certain cases on Fig. \ref{fig:1} as well as on Fig. \ref{fig:2}. For low values of $d_e$, a high and uniform current density allows the parallel alignment of a high density of magnetic field lines (see color map of $u_e$ for $d_e=0.05$ and $\rho_s=0$).
On the other hand, for high $d_e$ values, the current is not uniform enough along the layer for the magnetic field lines to line up perfectly, since their density decreases in the region where the current is weaker, (see $d_e=0.6$ and $\rho_s=0$). As we discuss below, this latter case is less likely to develop plasmoids. Finally, in the limit $\rho_s=0$, we obtain the approximate scalings $L_{cs} \propto d_e^{-1/10}$ and $A_{\rm{cs}} \propto d_e^{-1}$.

When $\rho_s$ is taken into account, ion sound Larmor effects can become important and the CS changes into a cross shaped structure aligned with the magnetic island separatrices \cite{Caf98}. Indeed, in Fig. \ref{fig:2}, when $\rho_s$ is increased (for $\rho_s \sim d_e$), a part at the end of the layer splits to extend along the separatrices (see $d_e=0.05$ and $\rho_s=0.05$). Here, the measured $L_{\rm{cs}}$ still corresponds to the length distributed symmetrically on both sides of $y=0$. We measured $L_{\rm{cs}} \propto \rho_s^{-1/2}$. As for the aspect ratios, they scale as $A_{\rm{cs}} \propto \rho_s^{-0.6}$ and $A_{\rm{outf}} \propto  \rho_s^{-1/2}$. For the series of simulations with $\Delta'_{\rm{box}}=14.3$ and $d_e=0.1$, the reconnection process occurs without forming any plasmoids (gridded red region) until $\rho_s \sim 0.4$.

For $\rho_s \gtrsim d_e$ and $\Delta'_{\rm{box}}=60$ (green diagonally striped region), the aspect ratio $A_{\rm{cs}}$ is sufficiently large and one plasmoid emerges from the center of the CS. This corresponds to a low wavenumber fluctuation that develops in the CS, which is entering the nonlinear phase.

\begin{figure}
    \centering
    % to crop the figures: trim=left bottom right top
\includegraphics[trim=0 0 0 0, scale=0.22]{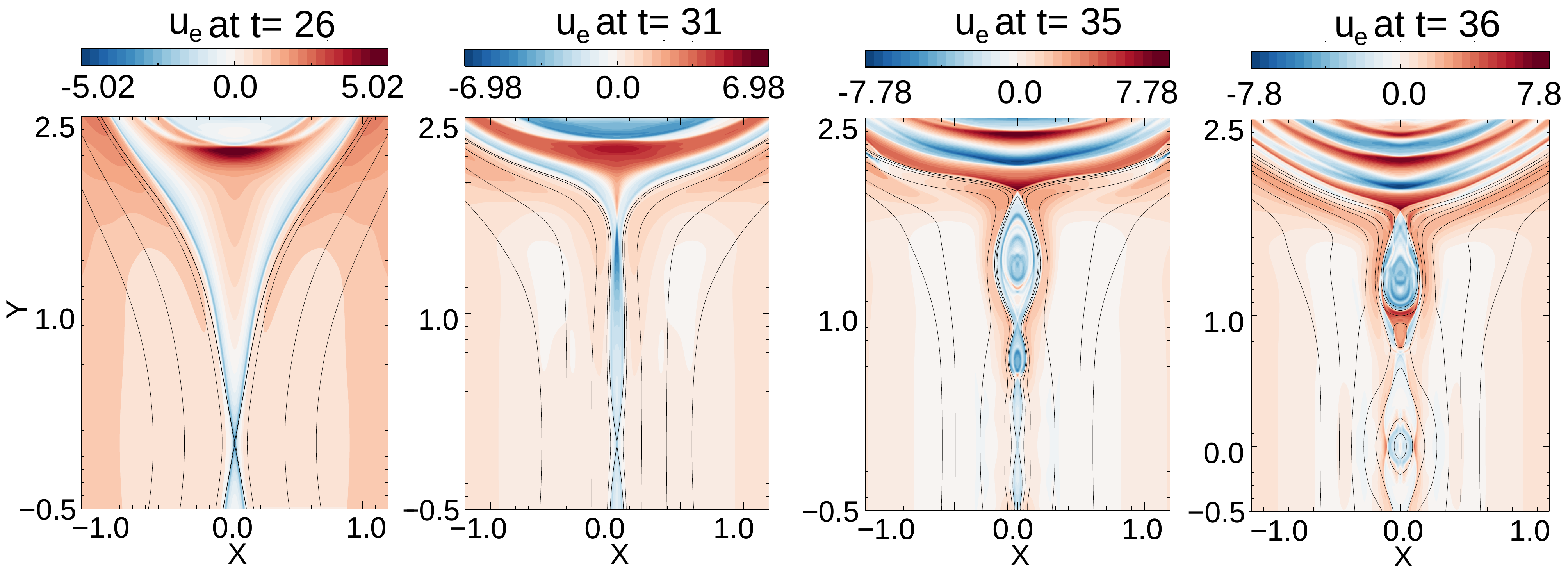}
\vspace{-0.4cm}
    \caption{Time evolution of the parallel current density. The plots show the color maps of $u_e$, while black lines are contour lines of $\apar$. For this simulation $d_e=0.085$, $\rho_s = 0.3$ and $\Delta'_{\rm{box}}=30$.}
    \label{fig:rhoslarge}
\end{figure}

For $\rho_s \gg d_e$ (green dotted region), the CS reaches a perfect cross shape \cite{Caf98}. This very different geometry can still lead to a more complex plasmoid formation. Indeed, in the regime $\rho_s \gg d_e$, the first plasmoids that break up the CS are symmetrically located above and below the $X$-point. This process is detailed in Fig. \ref{fig:rhoslarge}. We observe 4 main phases: (I) formation of the $X$-shaped current, (II) its ends meet to form a local $Y$-shaped CS, (III) plasmoids emerge and enter the nonlinear phase, (IV) they are expelled by the outflow and the center plasmoid emerges. This type of plasmoid onset takes place for $\rho_s > 0.4 \gg d_e$ in Fig. \ref{fig:2}.

\begin{figure}
    \centering
    % to crop the figures: trim=left bottom right top
\includegraphics[trim=10 0 0 0, scale=0.95]{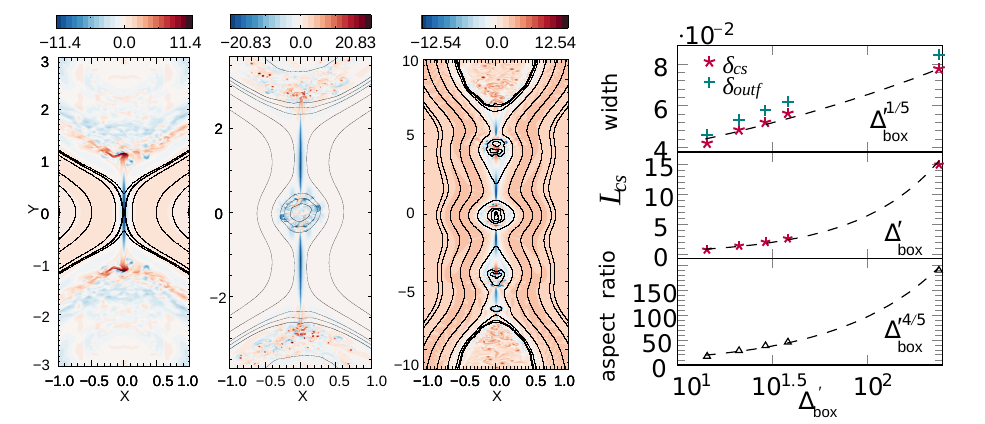}
\vspace{-0.45cm}
    \caption{CS characteristics for varying $\Delta'_{\rm{box}}$ at fixed $d_e=0.085$ and $\rho_s=0$. From left to right: color maps of $u_e$ with isolines of $\apar$ in black for $\Delta'_{\rm{box}}=14.3$, $\Delta'_{\rm{box}}=38$, $\Delta'_{\rm{box}}=240$ and plot of the same quantities as in Fig. \ref{fig:1} but here as functions of $\Delta'_{\rm{box}}$.}
    \label{fig:3}
\end{figure}

We now discuss the dependence on the $\Delta'_{\rm{box}}$ parameter, for $\rho_s=0$ and for $\rho_s \gg d_e$. In order to clearly identify a CS, we have considered large $\Delta'_{\rm{box}}$ values, which vary from $11.3$ to $240$.
For $\rho_s=0$ (Fig. \ref{fig:3}), the $L_{\rm{cs}}$ depends linearly on $\Delta'_{\rm{box}}$, as in the resistive case \cite{Wae89,Jem03,Jem04,Lou05}. We do not obtain plasmoids for $\Delta'_{\rm{box}} \leq 14.3$, in agreement with \cite{Del03} where this regime is shown to be prone to the development of the Kelvin-Helmholtz instability.
In the cases with $21 \leq \Delta'_{\rm{box}} \leq 38$, one plasmoid emerges and breaks up the reconnecting CS. For $\Delta'_{\rm{box}}=240$, two other plasmoids are formed when the reconnecting CS becomes more elongated (unstable) as $\Delta'_{\rm{box}}$ increases. In the limit $\rho_s = 0$, the outflow channel follows the CS and we observe indeed the scaling $A_{\rm{outf}} \propto \Delta'_{\rm{box}}$ (not shown here).

\begin{figure}
    \centering
    % to crop the figures: trim=left bottom right top
\includegraphics[trim=0 0 0 0, scale=1.1]{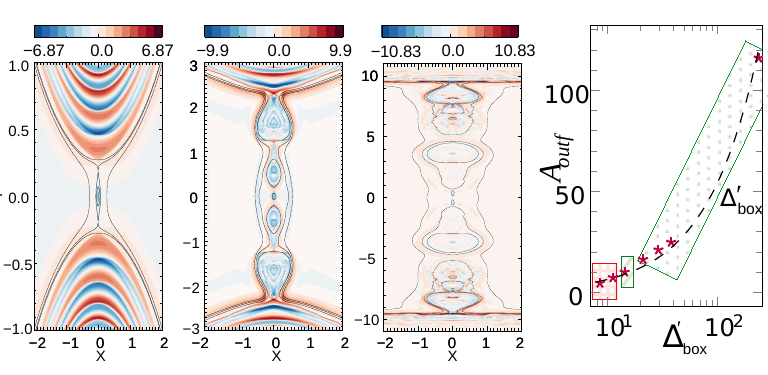}
\vspace{-0.55cm}
    \caption{CS characteristics for varying $\Delta'_{\rm{box}}$ at fixed $d_e=0.085$ and $\rho_s=0.3$. From left to right: color maps of $u_e$ with isolines of $\apar$ in black for simulations with $\Delta'_{\rm{box}}=14.3$, $\Delta'_{\rm{box}}=38$, $\Delta'_{\rm{box}}=240$, and aspect ratio of the outflow velocity channel before the onset of plasmoids. }
    \label{fig:4}
\end{figure}

For $\rho_s \gg d_e$ (Fig. \ref{fig:4}), on the other hand, the case with $\Delta'_{\rm{box}}=14.3$ is plasmoid unstable. In this regime, the small-scale, oscillating current layer pattern located inside the two magnetic islands, identified in Refs. \cite{Gra01, Del03}, is visible on the two left panels.
In the rightmost panel of Fig. \ref{fig:4}, we show the measured aspect ratio of the outflow velocity channel just before the appearance of the first plasmoid. For the least unstable reconnecting CS ($\Delta'_{\rm{box}}=14.3)$, we measured $L_{\rm{outf}}=1.21$ and $\delta_{\rm{outf}} = 0.12$, which implies a steady state reconnection rate of $R_{\rm{rec}} \sim (\delta_{\rm{outf}}/L_{\rm{outf}}) v_A B_{\rm{up}} \sim 0.1 v_A B_{\rm{up}}$. The red area corresponds to stable cases. The green striped area corresponds to the onset of only one plasmoid located at the center of the CS. Finally, the green dotted region corresponds to the cases where the first plasmoids emerge from a local $Y$-shaped CS (as described on Fig. \ref{fig:rhoslarge}).

\begin{figure}
    \centering
    % to crop the figures: trim=left bottom right top
\includegraphics[trim=0 0 0 0, width=9.5cm]{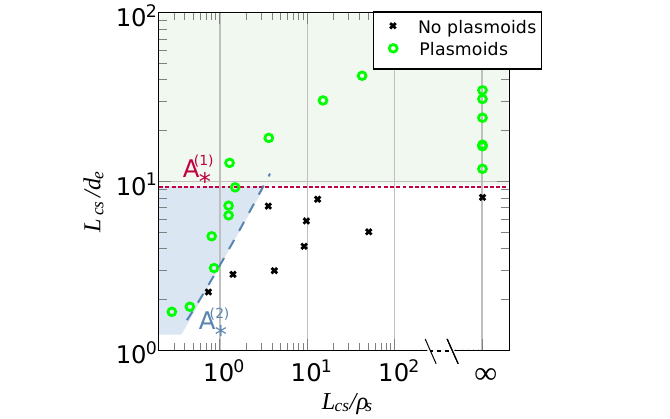}
\vspace{-0.35cm}
    \caption{Parameter space diagram that identifies where the plasmoid instability leads to the break up of the CS. The cases that give plasmoids (green dots) are either above the $A^{(1)}_\star$ or $A^{(2)}_\star$ thresholds. 
    %The green area is the region where $A_{\rm{cs}}$ is above the $A^{(1)}_\star$ threshold. The blue area, in the regime $d_e \ll \rho_s$, shows the region $A > A^{(2)}_\star$. The threshold $A^{(2)}_\star$ is.
     The threshold $A^{(2)}_\star$ is approximated by the scaling $L_{{\rm{cs}},\star}^{(2)}/d_e \propto L_{\rm{cs}}/\rho_s$ from Eq. (\ref{formula}). }
    \label{fig:diagram}
\end{figure}

We can construct a parameter space diagram (Fig. \ref{fig:diagram}), analogously to what was done for reconnection induced by plasma resistivity \cite{Ji11,Dau12, Cas13}, which allows one to identify the collisionless plasmoid regimes that take place once a reconnecting layer of a certain length is formed. According to our numerical simulations, the critical aspect ratio above which plasmoids break up the reconnecting CS is $A^{(1)}_\star=(L_{\rm{cs}}/\delta_{\rm{cs}})^{(1)}_\star \sim (L_{\rm{cs}}/d_e)^{(1)}_\star \sim 10$ when $d_e \gg \rho_s$. On the other hand, for $\rho_s \gtrsim d_e$, the plasmoid formation has a different threshold $A^{(2)}_\star =(L_{\rm{cs}}/\delta_{\rm{cs}})^{(2)}_\star$, and the simulations indicate $A^{(2)}_\star < A^{(1)}_\star$.

We can evaluate $A^{(2)}_\star$ by taking inspiration from the plasmoid instability theory presented in Refs. \cite{Com16,Com17,Hua17}. We consider a forming CS in which the amplitude of the tearing mode grows as $\apar(k,t) = A_0  \mbox{exp} ( \int_{t_0}^{t} \gamma (k)\mbox{d}t')$, where $\gamma$ and $k$ are the tearing mode growth rate and wavenumber, respectively, while $A_0$ is the magnetic flux amplitude at $t_0$. The plasmoid half-width is given by $w(k,t)= 2 (\apar a / B_{\rm{up}} )^{1/2}$, where $\apar$ is evaluated at the resonant surface and $a$ is the half-width of the CS.
We verified that the CS profiles are well fitted by a Harris sheet \cite{Biskamp2000}, for which $\Delta'_{\rm{cs}}=2 [(ka)^{-1} - ka]/a$. Given that the CS is slowly shrinking toward a finite width, we assume that, just before the plasmoid onset, the CS is in nearly steady-state and we neglect its time dependence. From Eqs. (\ref{fluid1})-(\ref{fluid2}), one can derive the dispersion relation of the collisionless tearing mode for arbitrary values of $\Delta'_{\rm{cs}}$ \cite{Por91}. For the marginally stable CS, one can consider the limit $\delta_{\rm{in}} \Delta'_{\rm{cs}} \ll 1$, with $\delta_{\rm{in}}$ indicating the width of the inner tearing layer. In this case, the full dispersion relation \cite{Por91} reduces to 
$\gamma^{(1)} = \left[ {\Gamma(1/4)}/{2 \pi \Gamma(3/4)}\right]^2 \Delta_{\rm{cs}}^{'2} d_e^3 k/a $
for $\rho_s^2  \ll d_e^2$, while $\gamma^{(2)} = {\Delta'_{\rm{cs}} d_e \rho_s k}/{a \pi}$ when $\rho_s^2 \gg d_e^2$.

We denote by ${\tau}_\star=L_{{\rm{cs}},\star}/ v_{A}$ the timescale for the plasma to be expelled from the CS because of the Alfv{\'e}nic outflow. If the magnetic flux amplitude becomes nonlinear (with plasmoid half-width $w_{\rm{nl}}$) in a time shorter than ${\tau}_\star$, the CS is broken by at least one plasmoid. Otherwise it remains stable. Therefore, taking 
$w(k, {\tau}_\star )= 2 \left({A_0 a}/{B_{\rm{up}}} \right)^{1/2} e^{\frac{1}{2} {\tau}_\star \gamma}$, the threshold for the plasmoid formation can be written as
\beq
{\tau}_\star \gamma = 2 \; \mbox{ln}\left[\frac{w_{\rm{nl}}}{2}\left( \frac{B_{\rm{up}}}{A_0 a}  \right)^{1/2}\right] \, .
\eeq
Assuming that the needed amplification factor of the magnetic flux perturbation is the same for the $\rho_s^2 \ll d_e^2$ and $\rho_s^2 \gg d_e^2$ cases, requiring ${\tau}_\star^{(2)} \gamma^{(2)} \sim {\tau}_\star^{(1)} \gamma^{(1)}$, making use of the numerical result $(L_{\rm{cs}}/d_e)^{(1)}_\star \sim 10$, and considering that for $(k a)^2 \ll 1$ we have  $\Delta'_{\rm{cs}} k \sim 1/a^2 \sim 1/d_e^2$, with $k \propto 1/L_{\rm{cs}}$, gives us the threshold condition
\beq
\frac{L_{{\rm{cs}}}}{d_e} = \frac{L_{{\rm{cs}},\star}^{(2)}}{d_e} \propto \frac{L_{\rm{cs}}}{\rho_s} \, . \label{formula}
\eeq
We identified a proportionality coefficient for which the proposed scaling, shown by the dashed blue line in Fig. \ref{fig:diagram}, correctly captures the plasmoid formation that occurs for significantly lower values of the CS aspect ratio when $L_{\rm{cs}}/\rho_s \lesssim 1$.

While the aspect ratio of the CS controls the plasmoid growth, the aspect ratio of the plasma flow channel regulates the rate of inflowing plasma via mass conservation. For $d_e \gg \rho_s$, the aspect ratios $A_{\rm{cs}}$ and $A_{\rm{outf}}$ essentially coincide since the plasma behaves as a one fluid. Therefore, for an incompressible flow in steady state, the marginal stability threshold $A^{(1)}_\star \sim 10$ yields the reconnection rate $R_{\rm{rec}} \sim 0.1 v_A B_{\rm{up}}$. On the other hand, for $\rho_s \gtrsim d_e$, two-fluid effects lead to a decoupling of the plasma flow channel from the electric current density, and in this case we find that $R_{\rm{rec}} \sim (\delta_{\rm{outf}}/L_{\rm{outf}})^{(2)}_\star v_A B_{\rm{up}} \sim 0.1 v_A B_{\rm{up}}$ even when $A^{(2)}_\star \ll  A^{(1)}_\star$. Since the global reconnection rate is controlled by the marginally stable CS \cite{ComBha16}, eventually $R_{\rm{rec}} \sim 0.1 v_A B_{\rm{up}}$ in the entire green and blue parameter space regions of Fig. \ref{fig:diagram}.

In summary, we have identified, with two-fluid numerical simulations and analytical arguments, the marginal stability conditions for the development of plasmoids in collisionless reconnecting CS. We find that in the collisionless regime, reconnecting CS are unstable to the formation of plasmoids for critical aspect ratios that can be as small as $L_{\rm{cs}}/\delta_{\rm{cs}} \leq 10$. For the marginally stable CS, we find that the aspect ratio of the outflow channel is $L_{\rm{outf}}/\delta_{\rm{outf}} \sim 10$ independent of the microscopic plasma parameters. The space of collisionless plasma parameters ($L_{\rm{cs}}/d_e$ and $L_{\rm{cs}}/\rho_s$) for which magnetic reconnection driven by electron inertia occurs in the plasmoid-mediated regime is organized in a new phase space diagram for collisionless reconnection. A new phase space diagram spanned by $L_{\rm{cs}}/d_e$ and $L_{\rm{cs}}/\rho_s$ for collisionless reconnection is presented, which extends the diagram of plasmoid onset for collisional plasmas \cite{Ji11}. Our results allow one to separate the collisionless laminar regime of reconnection from the collisionless plasmoid-mediated regime. The properties of the marginally stable CS obtained in this study contribute to the understanding of the rate of collisionless reconnection mediated by the plasmoid instability.

\vspace{0.50cm}

\begin{acknowledgments}
C.G. acknowledges the financial support provided by
the Universitée Franco-Italienne through the Vinci program. 
This work benefits from the support of the CNR contract DFM.AD003.261 (IGNITOR)-Del. CIPE n.79 del 07/08/2017 and of the grant DOE DESC0021254. 
The numerical simulations were performed using the EUROfusion high performance computer Marconi Fusion and Gallileo100 hosted at CINECA (project FUA35-FKMR and IsC86$\_$MR-EFLRA), and the computing facilities provided by Mesocentre SIGAMME hosted by Observatoire de la Côte d'Azur.
\end{acknowledgments}

%In needed, we can add supplemental material 
%\appendix
%\section{Appendixes}

\nocite{*}

\bibliography{Granier_plasmoids}% Produces the bibliography via BibTeX.

\end{document}